# Connecting Data Science and Qualitative Interview Insights through Sentiment Analysis to Assess Migrants' Emotion States Post-Settlement

Sarah Knudson
University of Saskatchewan
Saskatoon, SK, Canada
sarah.knudson@usask.ca

Srijita Sarkar
University of Saskatchewan
Saskatoon, SK, Canada
srs298@mail.usask.ca

Abhik Ray
Washington State University
Pullman, WA, USA
abhik.ray@wsu.edu

**ABSTRACT**

Large-scale survey research by social scientists offers general understandings of migrants' challenges and provides assessments of post-migration benchmarks like employment, obtention of educational credentials, and home ownership. Minimal research, however, probes the realm of emotions or "feeling states" in migration and settlement processes, and it is often approached through closed-ended survey questions that superficially assess feeling states. The evaluation of emotions in migration and settlement has been largely left to qualitative researchers using in-depth, interpretive methods like semi-structured interviewing. This approach also has major limitations, namely small sample sizes that capture limited geographic contexts, heavy time burdens analyzing data, and limits to analytic consistency given the nuances of qualitative data coding. Information about migrant emotion states, however, would be valuable to governments and NGOs to enable policy and program development tailored to migrant challenges and frustrations, and would thereby stimulate economic development through thriving migrant populations. In this paper, we present an interdisciplinary pilot project that offers a way through the methodological impasse by subjecting exhaustive qualitative interviews of migrants to sentiment analysis using the Python NLTK toolkit. We propose that data scientists can efficiently and accurately produce large-scale assessments of migrant feeling states through collaboration with social scientists.

## 1. INTRODUCTION

A substantial literature in the social sciences focuses on migrants' processes of settlement into new geographic contexts. Most often, research on migrant settlement is focused on generating understandings of why certain socio-demographic groups migrate [17], why they choose particular destinations [14], which key challenges they encounter during their migration and settlement [15], and how they fare as they integrate into their new context, as measured through post-migration benchmarks like employment, obtention of educational credentials, and home ownership [6]. Policy-oriented analyses also use these empirical findings to generate ideas for improving migration-focused policies and programs.

Methodologically speaking, much of the migration literature uses large-scale survey research with closed-ended questions to assess migrant processes and outcomes, both longitudinally and cross-sectionally (for example, the United States General Social Survey, the Canadian Census, and the World Values Survey). While large-scale survey research is valuable insofar as it can generate data on very large and statistically representative populations, can be administered efficiently and at lower cost than other research designs involving extensive interaction with respondents, and is higher on reliability than other social science research methods (namely qualitative approaches such as interviews), it also has major drawbacks. These include its overall superficiality and lack of contextualization when soliciting information or opinions, and its tendency not to probe respondents for more detailed rationales or nuances behind responses to typically straightforward questions. Survey research's shortcomings thus contribute to its lower validity as compared to qualitative interpretive approaches [1]. Further, the administration of large-scale surveys can be laborious and resource-intensive to coordinate—even if done online [11].

Because surveys are focused on short question-and-response exchanges as opposed to open-ended and more natural recounting of personal narratives or feelings, an additional weakness lies in their lack of texture when assessing emotions or "feeling states" as compared to qualitative research methods. For instance, Wave 6 of the World Values Survey (2010-2014) assesses self-reported happiness across multiple nations through a closed-ended survey question offering the following response categories: Very happy; Rather happy; Not very happy; Not at all happy; No answer [18]. Respondents have no opportunity to contextualize their responses.

We suggest that while survey-based migration data are highly valuable to academic, governmental and NGO audiences, collecting and analyzing data that generate deeper understandings of migrants' emotions or feeling states is also central to better governance and can contribute in two key ways. First, it can improve public service delivery through more accurately targeted policies and programs on national, state/province, and local levels by responding to migrant reports of their challenges, concerns, and sources of happiness or appreciation. Its secondary payoff comes through economic development, since economies are strengthened when migrant populations can integrate into their new contexts by finding meaningful employment and completing educational credentials that can lead to upward social mobility. But, despite the utility of studying feeling states in the migrant experience, looking at emotions is typically done through qualitative research that—like survey research—has major limitations. Although exhaustive biographical interviews of migrants are ideal for generating data on their feeling states [4], and achieve high levels of validity by asking migrants to explain their experiences in their own words, these interpretive approaches are time-intensive at collection and analysis phases, involve smaller samples, are not scalable to larger populations, and are both subjective and complex to analyze [3].

In light of time and resource constraints in government and policy-making environments that make in-depth qualitative analyses too laborious, and given the limitations of survey research in generating data on feeling states, researchers wanting insight into migrant feeling states seem to have reached a methodological impasse. However, we contend that there is a way through the impasse. Specifically, we propose subjecting exhaustive qualitative interviews to sentiment analysis as a means of efficiently and accurately producing large-scale assessments of migrant feeling states. Developments in data science are enabling compelling opportunities for collaboration with social scientists through triangulation of research methods that leverages the strengths of each methodological approach and improves the overall quality of migrant data available to government and policy-making entities [5, 13, 16].

## 2. OVERCOMING THE METHODOLOGICAL IMPASSE THROUGH USE OF SENTIMENT ANALYSIS

Opinion mining tools that enable automated content analysis are an established method in computer science, and have been used since the early 1980s. Their popularity and the flexibility of potential applications grew sixfold from their inception to the turn of the millennium [9], and they have continued to grow in popularity in recent years, alongside the boom in online textual content. Their growth as methodological tools has been accompanied by an increase in academic research into the opportunities they afford for understanding feeling states across large populations [11, 12]. Defined as a process of categorizing bodies of text to determine feelings, emotions and attitudes towards particular issues or objects [7], sentiment analysis aims to determine both the polarity of an individual's feeling state (i.e., positive, negative) as well as the strength of the polarity (i.e., strongly positive, mildly positive, weakly positive, neutral, etc.); in short, it elucidates the opinion of the individual who produced the text [11].

Sentiment analysis is rooted in sociological, psychological and anthropological theories positing that individuals' emotional evaluations of situations provide good general assessments of how they feel—both consciously and unconsciously—about situations, and how they might react to the given situations. As such, sentiment analysis has become a popular tool for marketing research [3, 13], but has also gained popularity as a means of addressing governance challenges. It has, for instance, been used by the U.S. government to detect and monitor spikes in negative sentiments directed at certain authorities or governmental bodies in online fora [13], and thus offers an opportunity to gather "early feedback" from citizens about issues, events, or authority figures that evoke strong emotional responses [11]. Politicians also use sentiment analysis to assess citizens' reactions to their stances on key issues, and sentiment analysis is increasingly used to interpret publics' responses to stressful events, crises, and situations of uncertainty [3].

Current applications of sentiment analysis in governance-related realms suggest that the method should be put to further use to tackle a range of migrant-focused governance challenges, including migrants' and non-migrants' feelings about contemporary migration crises and their policy and economic implications, reactions to experiences of racial and ethnic discrimination amongst and toward migrant communities, and migrants' evaluations of how they feel as they settle into their new contexts. Analyses focused on migrants, as well as comparative assessments of migrant and non-migrant populations, would be instrumental to efforts at promoting better governance. In addition to these compelling applications of the method, other strengths of sentiment analysis include its ability to quickly and inexpensively assess large bodies of text with minimal human labor, its high reliability and scalability, greater analytic consistency than qualitative approaches, and the unobtrusive nature of its assessment of feeling states. As an unobtrusive measure of emotions, it can assess sentiments despite individuals in the sample having never been questioned directly about their feelings and opinions [13].

The respective strengths of in-depth, qualitative research methods (namely biographical interviews) and sentiment analysis, when combined in a research endeavor, can therefore offer assessments of migrants' feeling states that are richly contextualized and founded in detailed accounts, but can also be quickly and rigorously assessed for their overarching patterns. Below, we outline a pilot project in which we subjected exhaustive qualitative interviews of migrants to sentiment analysis using the Python NLTK toolkit to test the viability of our mixed-methods approach.

### 2.1. A PILOT STUDY OF MIGRANT FEELING STATES IN THE CONTEXT OF SASKATOON, CANADA: DATA AND METHODS

Data from our sample were initially collected in Saskatoon, Canada, as part of a qualitative sociological study of transitions from adolescence to adulthood and processes of decision-making in the realms of education, work, finances, and relationships. Data collection occurred between October 2014 and March 2015. Thirty-six participants aged 18 to 32 were interviewed for an average of 58 minutes, using an open-ended, biographical research method intended to elicit detailed narratives of the young adults' lives and decision-making processes. Of the 36 participants, 19 were migrants to Saskatoon—either from out of province or out of country—and spent considerable time discussing their migrant and post-settlement experiences. Saskatoon is a small but growing city in Canada's Prairie region, with a rapidly expanding migrant population and the youngest average population of all Canadian cities. At the time of the data collection, Saskatoon's economy was strong and unemployment rates were low [2].

Interviews were digitally recorded and subsequently transcribed. Qualitative content analysis began with open data coding, in which an exhaustive list of concepts and categories was developed. Subsequent coding phases involved collapsing and refining categories to focus on the most salient themes and patterns in the data. We used NVivo, a qualitative data analysis program, to facilitate the coding process. Our initial qualitative analysis enabled us to identify three major reasons why the young adults had chosen to move to Saskatoon (to pursue educational opportunities, for overall improvement in quality of life, and because of existing social networks in the city). However, our findings did not offer any overarching understandings of the migrants' feeling states post-settlement, or any insight into how their feelings might be patterned along certain socio-demographic lines.

We, therefore, turned to sentiment analysis to see if we could glean a sense of emotion or feeling-state out of the interviews. The data—interviews from the 19 (im)migrants to Saskatoon— required a certain amount of pre-processing before we were able to perform the sentiment analysis on it. As a first step, we filtered out the text of the questions from the transcriptions. This left us with only the answers. Next, we filtered out answers that were less than 3 words. The intuition behind this was that shorter answers are usually in response to objective questions. We then used the NLTK library available as an Application Programming Interface [10]. The results of the sentiment analysis give us "neutral," "positive," and "negative."

While the above analysis is useful, it gives us a very broad picture of the overall emotional state of the interviewer. Moreover, many of the interviewees have "neutral" sentiment as classified by the sentiment analyzer. When the sentiment analyzer classifies a piece of text as "neutral," it is essentially because there is a lack of polarity in the entire text. However, this is often because of certain questions warranting an objective answer. In this case, our next option is to perform a more fine-grained analysis over the answers to the individual questions. Since each line in our dataset corresponds to an answer to a single question, we perform sentiment analysis on each individual answer. From this fine grained analysis, one can then answer questions like, "What does the person think about this particular topic?" One can also compare the sentiments expressed by different migrants in response to the same questions.

## 3. FINDINGS
### Table 1: International Migrants

| No. | Gender | Name | Place of Birth and Raised In | Why in Saskatoon | Sentiment |
|---|---|---|---|---|---|
| 1 | M | Mike | China | School | Neutral |
| 2 | F | Veronica | India | School | Neutral |
| 3 | M | Mario | India | School and now FT Job | Neutral |
| 4 | F | Jane | Philippines | School | Neutral |
| 5 | M | Peter | Ireland | School | Negative |
| 6 | M | Nick | Germany; brought up in Greece | School and Family | Neutral |
| 7 | F | Sarah | Singapore | School | Negative |
| 8 | F | Jane | Cameroon, Africa | School | Neutral |
| 9 | F | Amy | China | School | Neutral |
| 10 | F | Camellia | China | School | Neutral |
| 11 | F | Amira | Egypt | Work, Family, and School | Positive |

### Table 2: Inter-provincial migrants

| No. | Gender | Name | Place of Birth And Raised In | Why in Saskatoon | Sentiment |
|---|---|---|---|---|---|
| 1 | F | Bridget | Born in Bermuda; spent teens in Toronto and Montreal | School | Neutral |
| 2 | M | Patrick | Toronto (Immigrant Parents-1st gen. Canadians) | School | Neutral |
| 3 | F | Layla | New Brunswick | Job | Negative |
| 4 | M | Joe | BC | Job | Neutral |
| 5 | Unspe-cified | J.J. | Winnipeg; grew up in SK | Work and Family | Neutral |
| 6 | F | Ria | Winnipeg (1st gen. Canadian) | School | Negative |
| 7 | F | Laura | Winnipeg | Family | Neutral |
| 8 | F | Sam | Pakistan (1.5 gen.) | Family and School | Neutral |

From Tables 1 and 2 we see that for a few of the candidates the sentiment analyzer picks out the overall sentiments that are expressed by the interviewee. The interviewees in Table 1 are migrants from foreign countries. Out of 11 international migrants, 1 is positive, 2 are negative, and 8 are neutral. From Table 2, we see that out of the 8 inter-provincial migrants 0 are positive, 2 are negative, and the remaining 6 are neutral. While the majority of the responses are positive, the fact that some of the migrants express polarized sentiments validates the utility of using this analytical technique to determine the emotional state of migrants. The results that we get from analyzing individual answers are given in Table 3 show that individual lines (corresponding to individual answers) have sentiment values as well. By simply eyeballing the results, the reader will observe that in almost every case, the number of negative responses exceeds the positive.[i] Whether this allows us to conclude that migrants are mostly unhappy people requires further analysis from other dimensions. Being able to perform this fine-grained analysis allows us to stratify the answers (and by consequence the questions) based on topic as well as objectivity/subjectivity. As a next step we plan to get rid of answers to objective-style questions, and only analyze the subjective answers. This can be done by simply looking at the answers which have some sentiment value associated with them and discarding the rest. The fine grained analysis also gives us the opportunity to use yet another Natural Language Processing technique called "topic modeling" to automatically infer topics from text—thus reducing the need for human evaluation.

**Table 3: Counts of sentiments on individual lines**

| Name | NegCount | PosCount | NeutralCount | ErrorCount |
|---|---|---|---|---|
| Amira | 0 | 0 | 0 | 145 |
| Ria | 122 | 42 | 44 | 0 |
| Bridget | 79 | 27 | 40 | 0 |
| Nick | 33 | 24 | 41 | 0 |
| Amy | 113 | 45 | 59 | 0 |
| Jane (Cameroon) | 108 | 30 | 39 | 39 |
| Sarah | 35 | 13 | 29 | 0 |
| Layla | 91 | 28 | 50 | 0 |
| Veronica | 0 | 0 | 0 | 86 |
| J.J. | 50 | 35 | 62 | 0 |
| Laura | 0 | 0 | 0 | 211 |
| Mario | 0 | 0 | 0 | 204 |
| Patrick | 0 | 0 | 0 | 177 |
| Sam | 58 | 28 | 31 | 0 |
| Peter | 0 | 0 | 0 | 223 |
| Mike | 72 | 34 | 55 | 0 |
| Camellia | 102 | 47 | 75 | 0 |
| Jane (Philippines) | 162 | 51 | 93 | 0 |
| Joe | 0 | 0 | 0 | 171 |

## 3.1. FURTHER OPPORTUNITIES FOR TRIANGULATION IN MIGRANT RESEARCH USING SENTIMENT ANALYSIS

Sentiment analysis has thus shown itself to be a useful analytical tool that social scientists can use to efficiently discover sentiment from in-depth subjective interviews. While previously these interviews could only be done by a human analyst, automated sentiment analysis techniques greatly streamline this process.

Recent advances in Big Data analytical techniques have created many tools to automatically parse and analyze text, but the bottleneck in analyzing qualitative interviews remains in the interview design and data collection phase. Sentiment analysis of migrant interviews such as these could perhaps guide the creation of more subjective interviews as we understand which questions polarize migrants the most. In fact, this methodology could be applied to the design of qualitative interviews in any field.

A further consideration for alleviating the bottleneck at the interview design and data collection phase involves investing more effort in developing and testing interview methods that require lesser resource burdens from the research team, but that still succeed in collecting rich qualitative data. For instance, research questions could be sent to or accessed online by participants who would then "self-interview" using digital recording functions on their smartphones or tablets, and send their recordings to the research team for analysis. Granted, such an approach is limited by its lack of a live interviewer to probe respondents for more detailed answers—thereby compromising the richness of the data—and would exclude segments of populations with lower socioeconomic status who lack easy access to the required technology.

The results that sentiment analysis produces could be used to inform governments and policy makers/influencers to carefully assess the emotional states of migrants and thereby ensure a smoother migration/integration process. Governments could also use this information to encourage migration by quantifying the happiness of the new migrants. Further, governments could efficiently assess responses to potential changes in policy. To facilitate knowledge transfer between governmental and policy-making contexts and broader publics, sentiment analysis output can be quickly converted into graphic representations that are legible by citizens without specialized statistical or computer science training [7, 13].

The sentiment analysis method, as used in this paper, is somewhat limited in that it gives very coarse gradations, i.e., "positive," "negative," and "neutral." The technique can be upgraded to give further classifications such as "very happy," "somewhat happy," "very unhappy," etc. Also, the technique does not take into account the historical behavior of a person or sarcasm [8]. As Gaspar and colleagues argue [3], rigorous sentiment analyses ought to generate knowledge of feeling states beyond initial understandings of sentiment valences (e.g., "good" and "bad," "positive" and "negative") in texts. We need to delve deeper to understand the possible variations of context in which "good" and "bad" sentiments are voiced, and this can be achieved through progressively finer-grained analyses using topic modeling techniques. Coarser gradations and neutral results in sentiment analysis are also typical when qualitative interviews are based largely on asking more structured series of questions, as opposed to encouraging less structured, biographical and narrative-style accounts. Consequently, we also encourage interview designs that invite open-ended and detailed responses [4] to enable greater nuance in the data.

As tools for sentiment analysis and research into migrant feeling states continue to develop, we expect that the procedures we have outlined here can form a helpful basis for promoting better governance by generating deeper understandings of a broad range of migrant issues, both longitudinally and cross-sectionally, and of crisis situations as well as non-acute circumstances. And, although further triangulation of our methods with analyses of web-based comments for and about migrants is beyond the scope of this paper, we hope that researchers will do so in the near future.

## 4. CONCLUSION

We show the applicability of sentiment analysis as a tool to extract emotion from qualitative interviews. We apply it to the entire interview as well as to the individual responses to questions. We also propose using topic modeling along with sentiment analysis in order to perform a finer grained and more informed analysis. Our results show that for entire interviews the sentiments expressed are predominantly "neutral"; however, certain migrant interviews have shown a clear polarization. We require further analysis as well as more data before being able to significantly make any conclusions about the feeling states of migrants.

Sentiment analysis has shown itself to be a valuable tool as it can greatly hasten qualitative analysis. It took mere minutes for the program to discover what would have taken days or weeks for a human analyst. The tool is not infallible—in its current avatar the sentiment analyzer cannot provide nuanced sentiment classifications or detect sarcasm. These modifications along with performing a more rigorous analysis using various stratifications of the interview text are a part of our future work.

## 5. ACKNOWLEDGEMENTS
The authors wish to thank the Social Sciences and Humanities Research Council of Canada for the Insight Development Grant that enabled the initial qualitative data collection and analysis.

---

[i] In some cases, the analyzer returns Error values. Due to lack of time, we were unable to investigate and rectify this.